\documentclass[a4paper,10pt]{article}
	\usepackage{enumerate}
	\usepackage{color}
	\usepackage[utf8]{inputenc} 
	\usepackage[english]{babel}
	\usepackage[T1]{fontenc}
	\usepackage{graphicx}
	\usepackage{amsfonts,amssymb,amsmath,latexsym,amsthm}
	\usepackage{textcomp}
	\usepackage[pdftex]{hyperref}
	\usepackage{geometry}
	\usepackage{tikz}
	\newcommand{\C} {\mathcal{C}}
	\newcommand{\X}{\mathbf{ X}}
	\newcommand{\Y}{\mathbf{ Y}}

	\author{  M. V. Flamarion$^1$, T. Gao$^2$ , R. Ribeiro-Jr$^3$ \& A. Doak$^4$}
	\title{Flow structure beneath periodic waves with constant vorticity  under normal electric fields }

	\date{}
	
\begin{document}
	\maketitle
	\begin{center}
		
		{\footnotesize $^1$Unidade Acad{\^ e}mica do Cabo de Santo Agostinho, 
	UFRPE/Rural Federal University of Pernambuco, BR 101 Sul, Cabo de Santo Agostinho-PE, Brazil,  54503-900 \\
	marcelo.flamarion@ufrpe.br }

	\vspace{0.3cm}
	
	{\footnotesize $^{2}$ School of Computing and Mathematical Sciences, University of Greenwich, London SE10 9LS, UK.\\ t.gao@gre.ac.uk}

		\vspace{0.3cm}

		{\footnotesize $^{3}$ UFPR/Federal University of Paran\'a,  Departamento de Matem\'atica, Centro Polit\'ecnico, Jardim das Am\'ericas, Caixa Postal 19081, Curitiba, PR, 81531-980, Brazil \\ robertoribeiro@ufpr.br}
		
	\vspace{0.3cm}
		{\footnotesize $^{4}$ Department of Mathematical Sciences, University of Bath, Bath BA2 7AY, UK.\\ add49@bath.ac.uk }	
	

	\end{center}

	
	\begin{abstract} 
	\noindent Waves with constant vorticity and electrohydrodynamics flows are two topics in fluid dynamics that have attracted much attention from scientists for both the mathematical challenge and their industrial applications.  The coupling of electric fields and vorticity is of significant research interest. In this paper, we study the flow structure of steady periodic travelling waves with constant vorticity on a dielectric fluid under the effect of normal electric fields.  Through the conformal mapping technique combined with pseudo-spectral numerical methods, we develop an approach that allows us to conclude that the flow can have zero, two or three stagnation points according to variations in the voltage potential. We describe in detail the recirculation zones that emerge together with the stagnation points. Besides, we show that the number of local maxima of the pressure on the bottom boundary is intrinsically connected to the saddle points. 

		\end{abstract}

	\section{Introduction}
	
	The problem of surface water waves dates back to Stokes \cite{stokes1880theory}, who first considered surface gravity waves in deep water with a very large lengthscale in kilometres. The research on this subject has been extended in many directions. In particular, capillary waves, in which the surface tension is dominant  with a typical wavelength in millimetres, were studied first by Crapper \cite{crapper1957exact} where an exact solution was derived, and then by Kinnersley \cite{kinnersley1976exact} who generalised the results to the fluid sheet problem. When gravity and surface tension are equally important, it leads to a much more mathematical challenging problem named  capillary-gravity wave, with a small lengthscale about a  centimetre, which plays a vital role in transferring energy momentum and material fluxes across the ocean surface \cite{C,MK}.  A good monograph can be found in \cite{V} for a review.
	
	The aforementioned works assumed that the fluid is inviscid and the flow is incompressible and  irrotational. In practice, a theory with  non-zero vorticity is often more physically realistic. The most common choice on the vorticity is a constant value. A number of works have been achieved under such assumption for gravity waves. For example, Constantin and Strauss\cite{CS} conducted a rigorous analytical study. Thomas {\em et al.} \cite{TKM} investigated the modulational instability of such waves in the framework of a nonlinear Schr\"{o}dinger equation, and recently a more comprehensive work of stability analysis was achieved by  Francius and  Kharif \cite{FK}. Numerically speaking, a boundary integral equation method  was employed to compute fully nonlinear travelling-wave solutions with constant vorticity in this context by many authors \cite{SS,DST,V2,V3,V4}. In \cite{Ch}, time-dependent simulations were conducted by a hodograph transformation method, first pioneered by Dyachenko et al. \cite{D}, to examine the modulational instabilities numerically. 	In the presence of surface tension, i.e. capillary-gravity waves with constant vorticity, there have been fruitful achievements by different authors. For example,	steady-state solutions were computed by  Kang  and  Vanden-Broeck \cite{KV} via a boundary integral method.  A Nonlinear Schr\"{o}dinger Equation was derived in \cite{HKAC} to study the modulational instabilities. Time-dependent wave interactions in this context were investigated numerically in \cite{GMW} in which various travelling-wave solutions were also computed.
	
	All the studies in the last paragraph were interested solely in the shape of the free-surface, and integral properties such as wave energy and momentum. More recent studies  have considered the flow structure beneath the waves \cite{ribeiro2017flow,NR}. These structures are particularly interesting in the case of waves with constant vorticity. Of note,  stagnation points beneath the surface -- points within the bulk of the fluid that travels with the same speed as the wave. Such configurations can be also observed in waves with a point vortex \cite{gurevich1963vortex,shaw1972note,doak2017solitary,varholm2018stability}.  Another peculiarity of flows with constant vorticity is the emergence of pressure anomalies, such as  the maxima and minima of pressure on the bottom boundary being attained at locations  other  than beneath the crests and troughs of the free surface  respectively.
	
	In some practical situations is of interest to couple the fluid motion with electric fields. This subject of study is known as Electrohydrodynamics (EHD) and it can be applied widely in chemical engineering such as cooling systems in conducting pumps \cite{GM}, coating process \cite{GBMS} and etc. A good review paper was recently published by Papageorgiou \cite{P}.  Beginning with the early work and experiments of Taylor and McEwan \cite{TM}, researchers have been interested in the effect of electric fields on fluids with varying electrical properties. In that particular paper, normal electric fields were shown to be capable to destabilise the interface between two fluids. It was quickly followed by work on tangential electric fields by Melcher and  Schwarz \cite{MS}, where, on the contrary,  the interfacial waves can be stabilised due to such effects. Kelvin-Helmholtz and Rayleigh-Taylor instabilities were shown to be controlled and suppressed by tangential electric fields in \cite{ZK} and \cite{BPP,CPP} respectively.  Besides many works were achieved   theoretically by multi-scale technique, e.g. \cite{E,HV,GHPV,HVPP,PFE,PPV,LZW} in which weakly nonlinear model equations were derived for the two-dimensional problem. The readers are referred to the work by Wang \cite{Wang} for a comprehensive review. Travelling waves were computed by a boundary integral method, e.g. \cite{PV2,PV,DGVK,DGV}, for different configurations. Of note, in \cite{LZW,DGV}, the electric fields were shown to be capable of changing the type of the associated nonlinear Schr\"{o}dinger equation from focusing to defocusing such that dark solitary waves bifurcate at the phase speed minimum.     Nonlinear wave interactions in this context were studied theoretically and numerically by Gao et al. \cite{GWV}. Unsteady simulations were performed based on the time-dependent conformal mapping technique in  \cite{GDVW,GMPV} for a special case of capillary-gravity waves on a dielectric fluid of finite and infinite depth respectively covered above by a conducting gas layer. More recently, Gao et al. \cite{GWP} followed to investigate the singularity formation of the capillary-gravity wave due to the effect of normal electric fields.

	In this work, we consider a dielectric fluid in a two-dimensional space bounded below by an electrode, and above by a conducting passive gas, which in turn is bounded above by another electrode. A potential difference is applied between the electrode such that the electric field is imposed in the direction perpendicular to the undisturbed fluid surface. This configuration is well known for destabilising the interface between the fluids.  The fluid is assumed to be under the effect of constant vorticity. Our aim  is to study numerically the flow structure beneath waves with  constant vorticity under the influence of normal electric fields. More specifically, we  make use of the conformal mapping technique and pseudo-spectral numerical methods  to compute the locations of stagnation points and the pressure.  
	The rest of this paper is structured as follows. The mathematical formulation is given in section \ref{fm}.
	The linear  theory is studied in section \ref{ln}. The numerical scheme is presented in section \ref{ns}.  Section \ref{res}  is devoted to the results on the full nonlinear governing equations. Finally, concluding remarks are given in section \ref{conclusion}.

	\section{Mathematical Formulation}\label{fm}
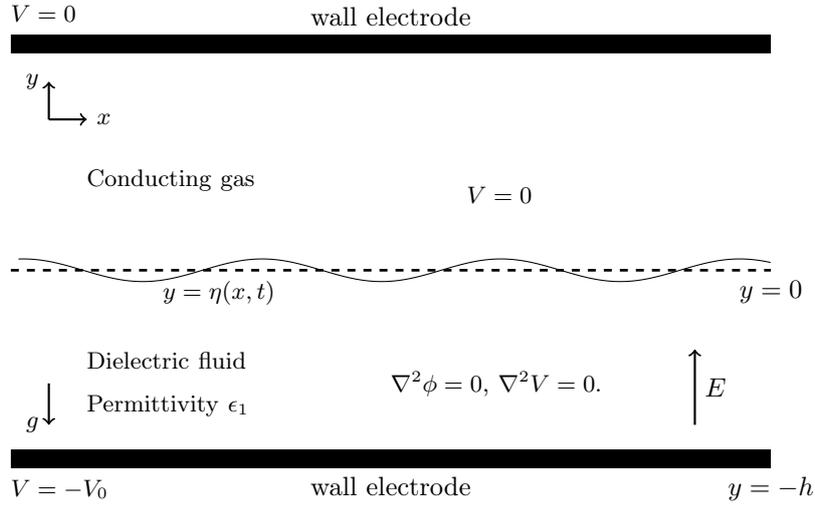
\begin{figure}
	    \centering
	    \begin{tikzpicture}
	\draw [ ->,thick] (9,-1.05) -- (9,-.05) node[right,midway]{$E$} ;
		\draw [ -,line width=.25cm] (0,4) -- (10,4) node[midway,above]{wall electrode};

	\draw [ -,line width=.25cm] (0,-1.5) -- (10,-1.5) node[left,below]{$y=-h$}node[midway,below]{wall electrode};
	\draw [ -,line width=1pt] (0,1) -- (10,1)[dashed] node[left,below]{$y=0$};
	\draw[thin]  plot [samples=300,domain=0.1:10] (\x,{1+.3*sin(2*(\x +pi/2-pi-.95) r)/2+0*sin((\x -pi+pi/2-.95) r)/2});
	\node[text width=2cm] at (7,2) {\small{$V=0$}};
	\node[text width=2cm] at (1,-1.9) {\small{$V=-V_0$}};
		\node[text width=2cm] at (1,4.4) {\small{$V=0$}};
	\node[text width=4cm] at (7,-.5) {\small{$\nabla^2 \phi=0$}, \small{$\nabla^2V=0$}.};
	\node[text width=4cm] at (3,-.2) {\small{Dielectric fluid}};
	\node[text width=4cm] at (3,-.8) {\small{Permittivity} $\epsilon_1$};
	\node[text width=4cm] at (3,2.2) {\small{Conducting gas}};
	\node [text width=2cm] at (3,0.7) {\small{$y=\eta (x,t)$}};

	\draw [ ->,thick]  (.5,-.5) -- (.5,-1.05)  node [left] {\small{$g$}};
	\draw [ ->,thick]  (.5,3) -- (.5,3.5)  node [left] {\small{$y$}};
	\draw [ ->,thick]  (.5,3) -- (1,3)  node [right] {\small{$x$}};
	
\end{tikzpicture}
	    \caption{Schematic of the problem.}
	    \label{fig:schm}
	\end{figure}

An inviscid and  incompressible dielectric fluid with density $\rho$ of mean depth $h$ and electric permittivity $\epsilon_1$ is considered in a two-dimensional space.  It is bounded below by a wall electrode and surrounded by a region occupied by perfectly conducting gas and enclosed by another wall electrode atop. In the electrostatic limit of Maxwell's equations, the induced magnetic fields are negligible so the electric field is irrotational due to Faraday's law. A potential function $V$ of the electric fields $\vec{E}$ can be then introduced such that $\vec{E}=\nabla V$. It follows that $V$ satisfies the Laplace Equation in the dielectric fluid layer. The potential difference between the two electrodes is denoted by $V_0$. The voltage potential is invariant in the conducting gas layer. Without losing generality, we set $V=-V_0$ on the bottom boundary and $V=0$ in the gas layer. A schematic of the problem is presented in Figure \ref{fig:schm}\,. The gravitational acceleration and the surface tension coefficient are denoted by $g$ and $T$, respectively. A Cartesian $x-y$  coordinate system is introduced with the gravity pointing in the negative $y$-direction and $y=0$ being the undisturbed free-surface. It follows that the bottom boundary of the fluid is at $y=-h$, and the upper boundary is free to move and denoted by $\eta(x,t)$. We consider a periodic wavetrain propagating at a constant speed in the positive $x$-direction. Also, we let $x=0$ be at a wave crest such that the wave profile is symmetric by the $y$-axis.
	
	The velocity field is assumed to be an irrotational perturbation of a shear flow, namely
	\begin{equation}\label{eq:velocity_field}
		\vec{U} = \hat{u} + \vec{u},
	\end{equation}
	where the vorticity is constant that equals $\gamma$, and $\hat{u} = \left(-\gamma(y+b),0\right)$ with $b$ being another constant. The velocity field  $\vec{u}$ is irrotational so that there exists a potential function $\phi$ satisfying $\vec{u}=\nabla \phi$. We seek steady waves of wavelength $\lambda$ in a frame of reference moving with the wave speed $c$. 	This is achieved by a change of variables: $X=x-ct$,  $Y = y$. It immediately follows that $\eta(x,t)=\eta(X)$. The fluid motion can be characterised by the potential function $\phi$ and the voltage potential $V$ 	as follows
	\begin{align}
		\nabla^2 \phi &= 0 & \text{in}\,\,\, -1<Y<\eta(X), \label{eq:field1}\\
		\nabla^2 V &= 0 &  \text{in}\,\,\, -1<Y<\eta(X), \label{eq:field2} \\
		- c \eta_X +  \left(\phi_X - \gamma\left(\eta + b \right)  \right) \eta_X &=\phi_Y & \text{for} \,\,\, Y=\eta(X), \label{eq:KBC2} \\
		\phi_Y &=0 & \text{for} \,\,\, Y=-1, \label{eq:KBC1}\\
		V&=0 & \text{for} \,\,\, Y=\eta(X), \label{eq:KBCV1}\\
		V&=-1 & \text{for} \,\,\, Y=-1, \label{eq:KBCV2}
	\end{align}
	where we have chosen $h$, $\sqrt{h/g}$ and $V_0$ as the reference length, time and voltage potential, respectively.
	Furthermore, the continuity of pressure on the free-surface yields the dynamic boundary condition, which can be written as
	\begin{equation}\label{eq:Bern1}
		-c\phi_X + \frac{1}{2} (\phi_X^2 + \phi_Y^2) + \eta - \gamma \left(\eta + b\right) \phi_X  + \gamma \psi - \tau \frac{\eta_{XX}}{\left(1+\eta_X^2\right)^{3/2}} + \frac{E_b}{2} \lvert \nabla V \rvert^2 =B. 
	\end{equation}
	Here, $\psi$ is the harmonic conjugate of $\phi$. Parameter $\tau$ and $E_b$ are the non-dimensional Bond and Electric Bond numbers respectively, given by
	\begin{equation}
		\tau = \frac{T}{\rho g h^2}, \hspace{1cm} E_b = \frac{\epsilon_1 V_0^2}{\rho g h^3}.
	\end{equation}
The hydrodynamic pressure in fluid body is computed by the Bernoulli equation 
	\begin{equation}\label{eq:Pressure}
		p = - \left(-c\phi_X + \frac{1}{2} (\phi_X^2 + \phi_Y^2) + Y - \gamma \left(Y + b\right) \phi_X  + \gamma \psi -B  \right), 
	\end{equation}
	and the stream function can be written as
	\begin{equation}\label{funcao de corrente}
		\psi_s(X,Y):=\psi(X,Y) - \gamma Y \left(\frac{1}{2}Y+b\right)  -cY.
	\end{equation}

	\section{Linear theory}\label{ln}
	
		\begin{figure}[!hb]
		\centering
		\includegraphics[scale=0.95]{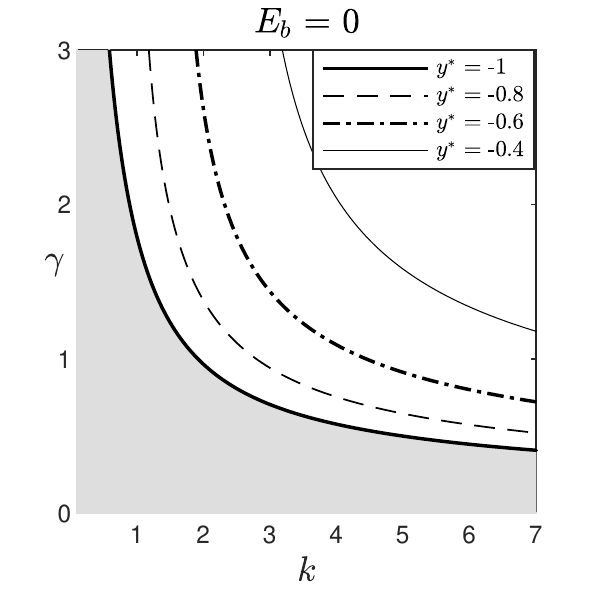}
		\includegraphics[scale=0.95]{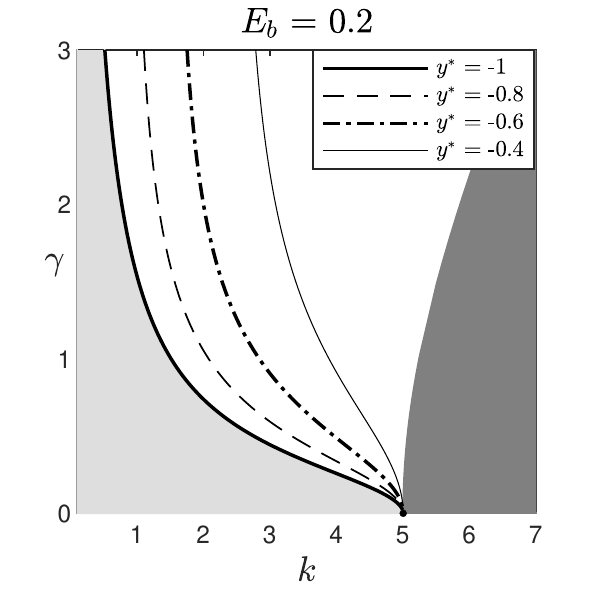}
		\includegraphics[scale=0.95]{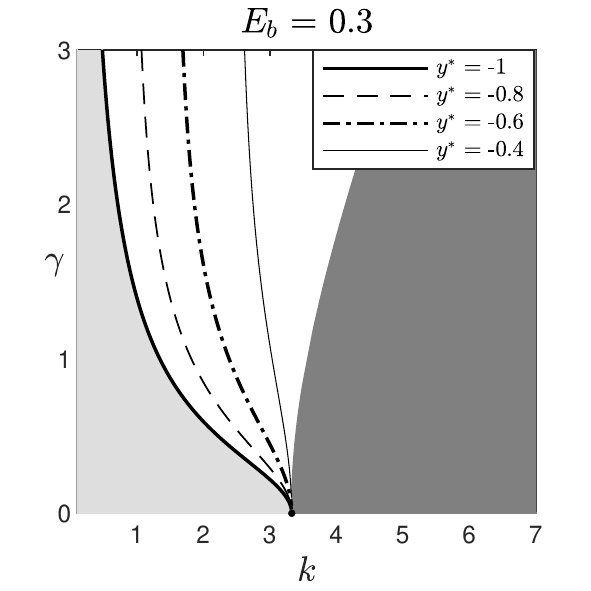}
		\includegraphics[scale=0.95]{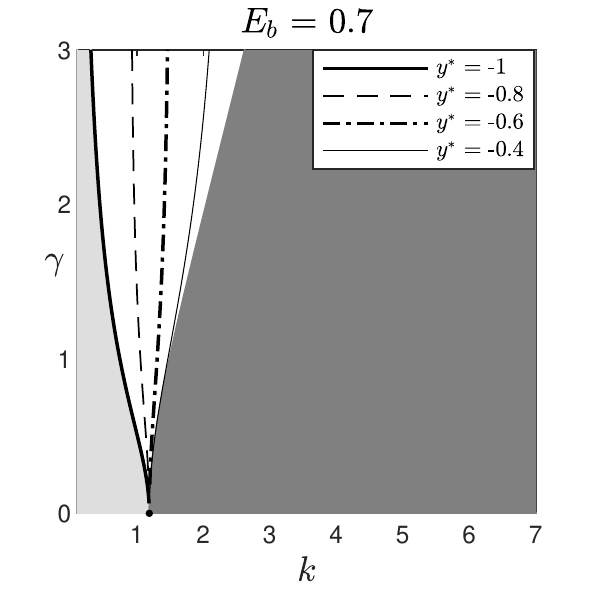}
		\caption{Parameter dependence of the linear flow in terms of the vorticity ($\gamma$) and the wave number ($k$) for different values of $E_b$. Flows without surface tension, i.e., $\tau = 0$.  }
		\label{fig:gamma_Vs_k1}
	\end{figure}

	In this section, a linear theory is developed from the governing equations. A trivial solution takes the form of
	\begin{equation}
		\phi_0(X,Y) = 0, \hspace{0.6cm} V_0(X,Y) = Y, \hspace{0.6cm} \eta_0(X)=0,
	\end{equation}
which is perturbed by a small disturbance, namely
	\begin{eqnarray}
		\begin{cases}\label{eq:linear1}
			\eta(X)=\epsilon \hat{\eta}, \\
			\phi(X,Y)=\epsilon \hat{\phi}, \\
			V(X,Y)= Y+\epsilon \hat{V},
		\end{cases}
	\end{eqnarray}
	in which $\epsilon$ is a small parameter that measures the wave amplitude. 	Solving the Laplace equations \eqref{eq:field1} and \eqref{eq:field2} together with the boundary conditions \eqref{eq:KBC1}--\eqref{eq:KBCV2} yields
	\begin{eqnarray}
		\begin{cases}\label{eq:linear2}
			\hat{\eta}(X)= \Re\left\{ A e^{ikX}\right \}\,, \\
			\hat{\phi}(X,Y)=\Re \left\{ B e^{ikX}\cosh k(Y+1)\right \}\,, \\
			\hat{V}(X,Y)= \Re\left\{ D e^{ikX}\sinh k(Y+1)\right \}\,.
		\end{cases}
	\end{eqnarray}
	in which $A$, $B$ and $D$ are unknown constants, and $k=2\pi/\lambda$ is the wavenumber. By dropping all the nonlinear terms in the kinematic condition \eqref{eq:KBC1} and the dynamics boundary condition \eqref{eq:Bern1}, the equations are linearised as
	\begin{eqnarray}
		&&-\left(c + \gamma b\right) \hat{\eta_X} = \hat{\phi_Y}\,, \label{eq:KBC2lin} \\
		&&\hat{V} = -\hat{\eta}\,, \label{eq:KBCV2lin}\\
		&&-\left(c + \gamma b \right)\hat{\phi}_X + \hat{\eta}  + \gamma \hat{\psi}  -  \tau \hat{\eta}_{XX} + E_b \hat{V}_Y = 0\,. \label{eq:Bern2}
	\end{eqnarray}
	Substituting (\ref{eq:linear1})--\eqref{eq:linear2} into equations (\ref{eq:KBC2lin} - \ref{eq:Bern2}), it is discovered that 
	\begin{align}
		B & = \frac{-iA(c+\gamma b)}{\sinh k }\,, \\
		D & = \frac{-A}{\sinh k }\,,
	\end{align}
and the linear dispersion relation:
	\begin{equation}\label{eq:c}
		c_{\pm} = -\gamma b + \frac{1}{2k} \left[ \gamma \tanh k \pm \sqrt{\Delta}        \right],
	\end{equation}
	where 
\begin{eqnarray}\label{eq:delta}
    \Delta = \gamma^2 \tanh^2 k +  4k\tanh k\left(1+\tau k^2 \right) - 4k^2 E_b\,.
\end{eqnarray}
As remarked in \cite{GMW}, $c_-$ and $c_+$ can be swapped by letting $\gamma\rightarrow-\gamma$, and hence it is sufficient to only consider $c_+$, i.e., the right-moving waves, hereafter.

	We follow to study the effect of the electric field on the position of the stagnation points in the fluid.
	The velocity field $\vec{U} = (U_1,U_2)$ is derived by (\ref{eq:velocity_field}) and (\ref{eq:linear1})  and takes the form of
	\begin{align}
	&	U_1 = -\gamma (Y+b)  + \frac{k\left(c+\gamma b\right) \epsilon A }{\sinh k} \left[\cosh(k(Y+1)) \cos(kx)  \right],\label{eq:linU}\\
	&	U_2 = \frac{k\left(c+\gamma b\right) \epsilon A }{\sinh k} \sinh(k(y+1)) \sin(kx). \label{eq:linV}
	\end{align}
	In the frame moving with the wave speed, a fluid particle  has the trajectory  $(X(t),Y(t))$ satisfying 
	\begin{equation}\label{ODE}
		\left\{
		\begin{array}{l }
			\dfrac{dX}{dt} = \phi_X(X,Y)  -\gamma(Y+b)   - c , \\  \noalign{\bigskip}
			\dfrac{dY}{dt} =  \phi_Y(X,Y).   \\  \noalign{\bigskip}
		\end{array} \right.
	\end{equation}
	Stagnation points are defined as the equilibrium points of the dynamical system \eqref{ODE}. We fix $b = 1$ in the remaining of the work.  We make this choice because it seems more natural consider a current with zero velocity on the bottom boundary.  Besides, note that the value of $b$ has no effect on the location of the stagnation points -- this can be seen from equations (\ref{eq:c}) and (\ref{ODE}). 
	
The equation (\ref{eq:linV}) shows that a stagnation point, if it exists, must occur at $x= n \lambda/2$, where $n\in \mathbb{Z}$.
The conditions for the existence of a stagnation point are given by
	\begin{equation}\label{eq:TDE1}
		\gamma Y = 
		\begin{cases}
			\hat{A} \left(c+\gamma b\right) \cosh(k(Y+1)) - (c + \gamma b)\,,  & x=n\lambda \,,\\
			-\hat{A} \left(c+\gamma b\right) \cosh(k(y+1))  - (c + \gamma b)\,, & x=\frac{(2n-1)}{2}\lambda \,,
		\end{cases}
	\end{equation}
	where $\hat{A} = \epsilon A k/\sinh k$. These transcendental equations can  be only solved numerically.

\begin{figure}[!ht]
		\centering
		\includegraphics[scale=1]{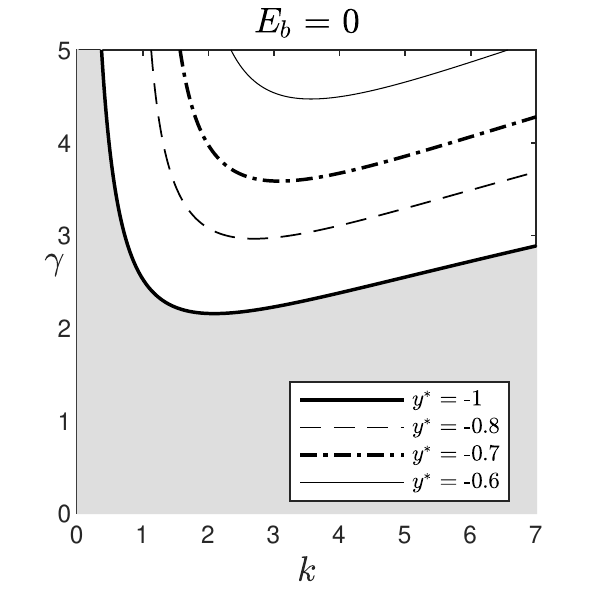}
		\includegraphics[scale=1]{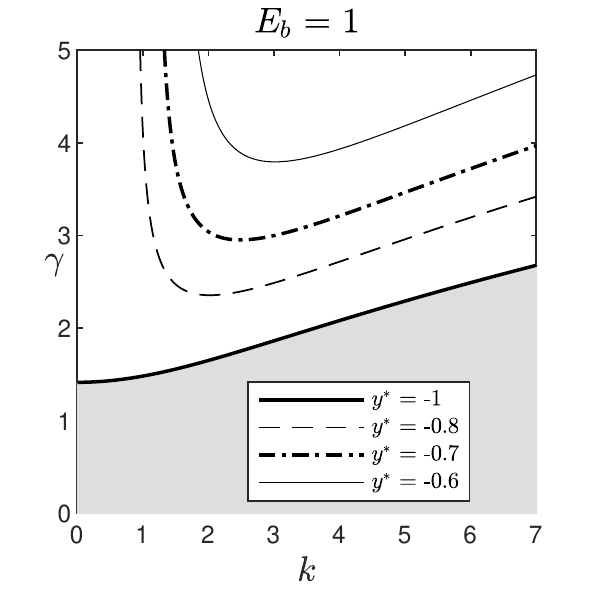}
		\includegraphics[scale=1]{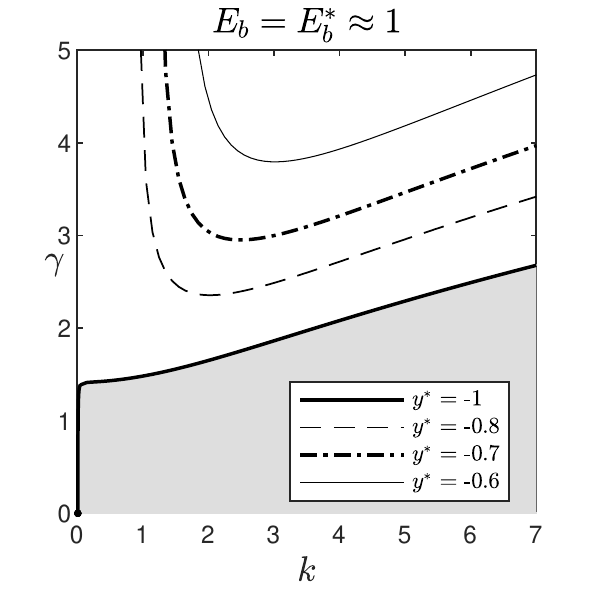}
		\includegraphics[scale=1]{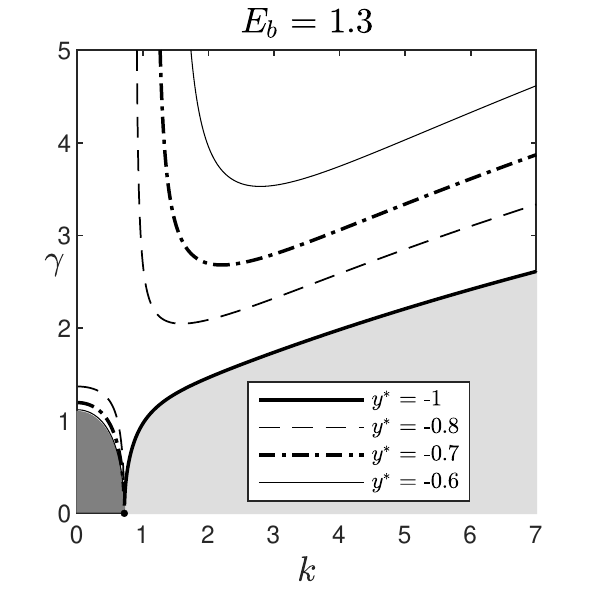}
		\includegraphics[scale=1]{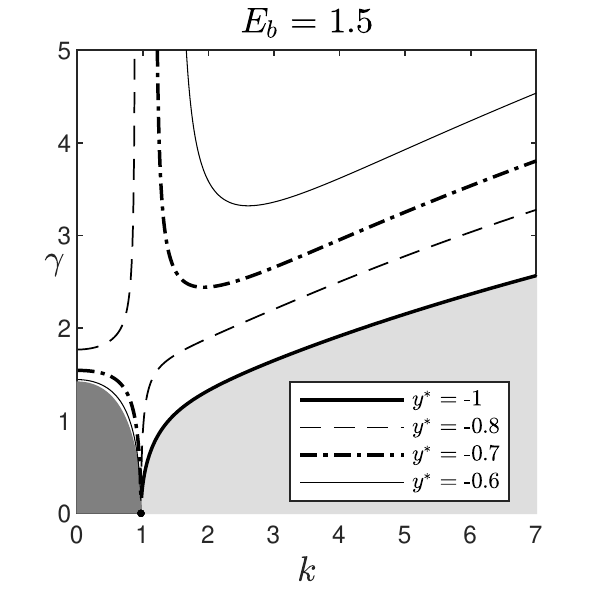}
		\includegraphics[scale=1]{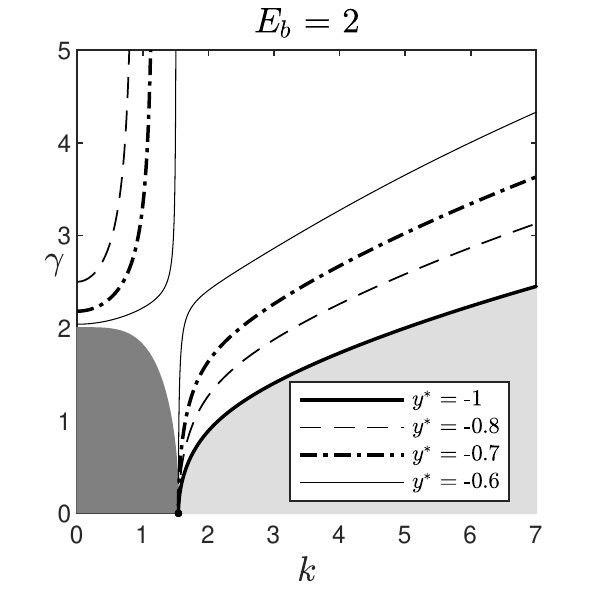}
		
		\caption{Parameter dependence of the linear flow in terms of the vorticity ($\gamma$) and the wave number ($k$) for different values of $E_b$. Flows with surface tension  $\tau = 1$.  
		}
		\label{fig:gamma_Vs_k2}
	\end{figure}

	Figure \ref{fig:gamma_Vs_k1} and  \ref{fig:gamma_Vs_k2}   show the $(k,\gamma)$-solution space of the linear flow for different values of $E_b$ when $\tau = 0$ and  $\tau = 1$, respectively.  The solution space is divided in three parts: (i) a region where there is no solution (the darker gray); (ii) a region in which there is no stagnation point in the flow (the lighter gray); (iii) a region in which there exist stagnation points beneath the linear waves (white region) -- the lines in this region are depicted to show the depth of a stagnation point denoted by $y^*$. As it can be seen from the figures, the increase of the parameter $E_b$ leads to the appearance of a region where there is no solution. In particular, for the case where $\tau = 1$,  there exist a threshold value of  $E_b$  such that long waves only exist for  $\gamma$ being sufficiently large.   Regarding the location of the  stagnation points, it is observed that the  depth of the stagnation points  decrease as  $\gamma$ is increased at a fixed wavenumber.  It is worth mentioning that there exist two different waves with stagnation points at the same depth for a fixed vorticity due to the presence of surface tension.

	\section{Numerical scheme}\label{ns}
	
	The numerical method employed to compute solutions of the equations (\ref{eq:field1})-(\ref{eq:Bern1})  is based on a conformal mapping  approach which transforms the free boundary conditions   to another one with a  simpler geometry. Following Ribeiro et al. \cite{ribeiro2017flow}, the physical domain is mapped conformally onto a horizontal strip of uniform thickness denoted by $D$. All the computations are conducted in the canonical domain. Here we present just a brief summary of the numerical scheme. 
	
	We denote by $(\xi,\zeta)$ the coordinates in the canonical domain and  by  $X(\xi,\zeta)$ and  $Y(\xi,\zeta)$  the horizontal and the vertical  components of the conformal map. It follows  that 
	\begin{equation}\label{eq:Y}
		Y(\xi,\zeta) = \mathbf{F}^{-1} \left[  \frac{\sinh(k_j(\zeta+D))}{\sinh(k_jD)}  \mathbf{F}_{k_j}[\mathbf{Y}]  \right] + \frac{\zeta}{D},  
	\end{equation}
	and
	\begin{equation}\label{eq:X}
		X(\xi,\zeta) =  \xi - \mathbf{F}^{-1} \left[i  \frac{\cosh(k_j(\zeta+D))}{\sinh(k_jD)}  \mathbf{F}_{k_j}[\mathbf{Y}]  \right], \quad j \neq 0 
	\end{equation}
	where $\mathbf{Y}(\xi) = \eta(X(\xi))$, $k_j = 2\pi j/L$, for $j\in \mathbb{Z}$, and
	
	$$ \mathbf{F}_{k_j}[g(\xi)] = \hat{g}(k_j) =  \frac{1}{L}\int_{-L/2}^{L/2}g(\xi) e^{-ik_j\xi}\, d\xi, $$
	$$  \mathbf{F}^{-1}\left[ \{\hat{g}(k_j)\}_{j \in \mathbf{Z}}\right] = \sum_{j = -\infty}^{+\infty} \hat{g}(k_j) e^{ik_j\xi}.$$
Here $[-L/2,L/2)$ is the computational domain and $L$ is chosen so that the wavelength in the  physical and in the canonical domain are the same. Thus,  $L = \lambda$ which yields the equation

	\begin{equation}\label{eq:D}
		D = \left< \mathbf{Y} \right> + 1,
	\end{equation}
	where 
	\begin{equation}
	\left< \mathbf{Y} \right> =  \frac{1}{L}\int_{-L/2}^{L/2} \mathbf{Y} (\xi)\, d\xi.
	\end{equation}

	The governing equations (\ref{eq:field1})-(\ref{eq:Bern1}) are written in the canonical coordinates, which  after some manipulations yields the following equation for the free-surface elevation  $\mathbf{Y}(\xi) = Y(\xi,0)$
	\begin{equation}\label{eq:Free surface}
		\begin{split}
			-\frac{c^2}{2} &- \frac{c^2}{2J}  + \mathbf{Y}+ \gamma^2 \frac{ (\C[(\mathbf{Y}+b)\mathbf{Y}_\xi])^2 - \left((\mathbf{Y}+b)\mathbf{Y}_\xi \right)^2}{2|J|} + \gamma^2\mathbf{Y}\left(\frac{\mathbf{Y}}{2} + b\right) \\
			&+ \frac{\left(c+\gamma \C [(\mathbf{Y}+b)\mathbf{Y}_\xi] \right) \left( c+ \gamma(\mathbf{Y}+b)(1 - \C[\mathbf{Y}_{\xi}]) \right)}{J} -\gamma c b  \\
			& - \sigma\frac{\X_\xi \Y_{\xi\xi} - \Y_\xi \X_{\xi\xi}}{J^{3/2}} + \frac{E_b}{2D^2J} = B,
		\end{split}
	\end{equation}
	where  $J(\xi) = \X_\xi^2 + \Y_\xi^2$ and $\X_\xi$ is the $\xi$-derivative of $\X(\xi) = X(\xi,0)$,  which is given by 
	\begin{equation}\label{eq:X_xi}
		\X_{\xi} =  1 - \C[\Y_{\xi}],
	\end{equation}
	with $\C[\cdot]:=\mathbf{F^{-1}}\left[ i\coth(k_jD)\mathbf{F}_{k_j}\left[\cdot\right] \right]$.
	
	As the free surface wave is an even function with crest at $X=0$,  we assume that  in canonical coordinates its  crest is located at $\xi = 0$. So we have the equation
	\begin{equation}\label{eq:height}
		\Y(0)- \Y(-L/2) = H,
	\end{equation} 
	where $H$ is the height of the wave. 
	 Moreover, we set the mean level of the free surface wave to be zero in the physical domain, which yields the equation
	\begin{equation}\label{eq:mean0}
		\int_{-L/2}^{0} \Y\X_\xi \, d\xi =0.
	\end{equation}

 Equations (\ref{eq:D}), (\ref{eq:Free surface}), (\ref{eq:height}) and (\ref{eq:mean0}) form a system of four equations and four unknowns: $\Y$, $c$, $D$ and $B$.    This system is discretised in $\xi$ by introducing collocation points uniformly distributed along $\xi$.
 By evaluating all the $\xi$-derivatives via a Fourier spectral method using the FFT, the partial differential equations recast as algebraic equations which can be solved by Newton's method as in \cite{ribeiro2017flow}.
	
	\begin{figure}[h!]
		\centering
		\includegraphics[scale=1]{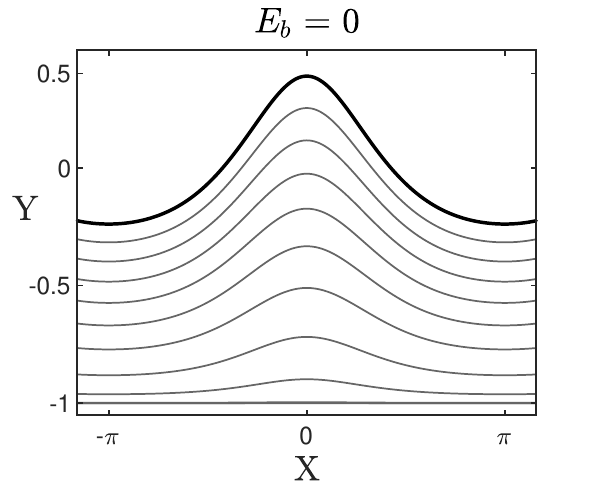}
		\includegraphics[scale=1]{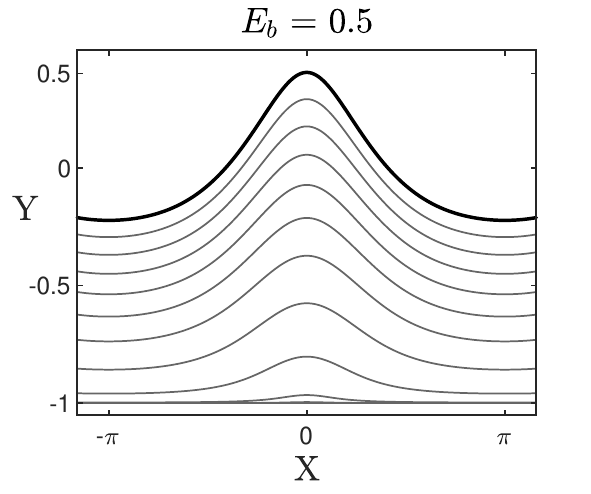}
		\includegraphics[scale=1]{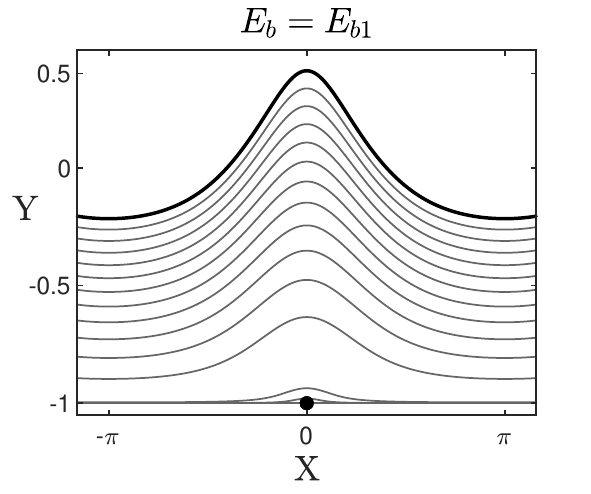}
		\includegraphics[scale=1]{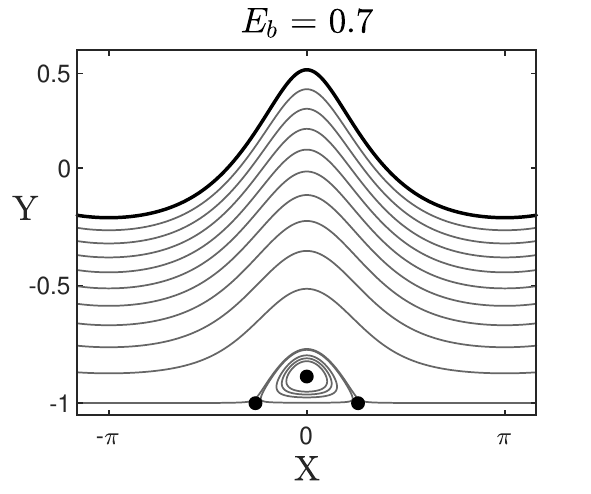}
		\includegraphics[scale=1]{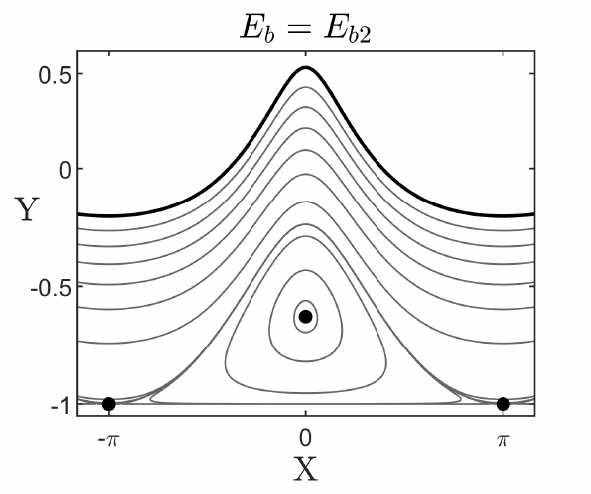}
		\includegraphics[scale=1]{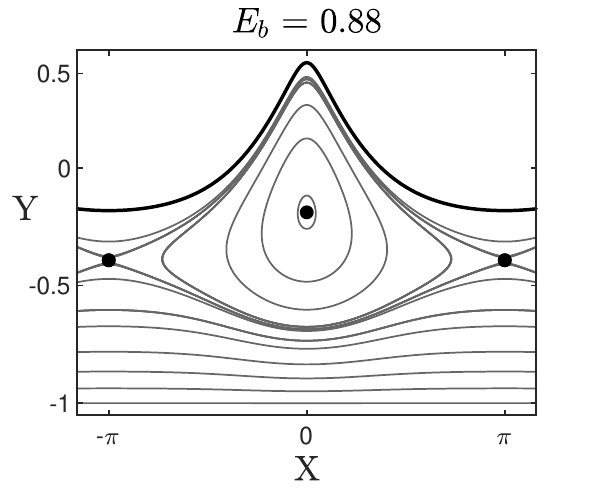}
		\caption{Phase portrait for different values of the  electric bond number  $E_b$ ($E_{b1} \approx0.638$ and $E_{b2} \approx0.803$). Parameters of the flow: $\tau = 0$ and  $\gamma = 0.5$.  }
		\label{fig:PhasePortraitTau_0}
	\end{figure}

	\section{Numerical results}\label{res}
		Ribeiro-Jr \emph{et al.} \cite{ribeiro2017flow} showed numerically that  the flow beneath a nonlinear periodic gravity wave on an inviscid fluid with constant vorticity 
	can have zero, one, two or three stagnation points. 	Besides, these authors discovered that the pressure on the bottom boundary has two points of local maxima  when the vorticity is sufficiently strong.  In this section, we revisit this problem with the addition of normal electric fields. For a better comprehension, we first investigate the effects of the electric Bond number  $E_b$ in the appearance of the stagnation points and the streamlines. Then, we discuss some pressure anomalies that arises in electric-capillary-gravity flows.  We consider  nonlinear waves   in a intermediate-depth regime. Thus, in all simulations presented   the	
		 typical wavelength is selected to be $2\pi$, namely $\lambda=2L=2\pi$, and  the wave steepness ($H/\lambda$) is $0.1$.

	\begin{figure}[h!]
		\centering
		\includegraphics[scale=1]{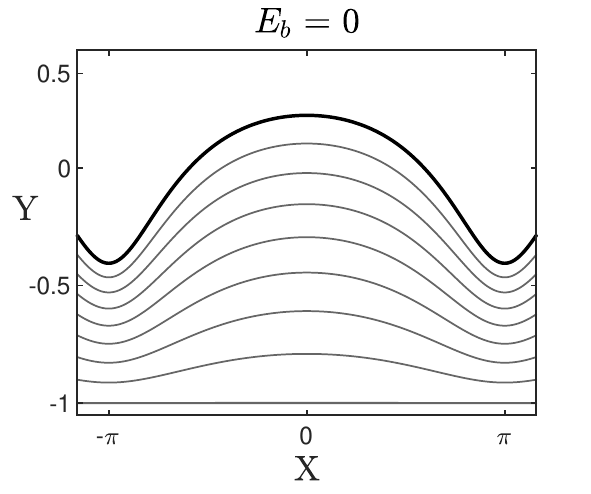}
		\includegraphics[scale=1]{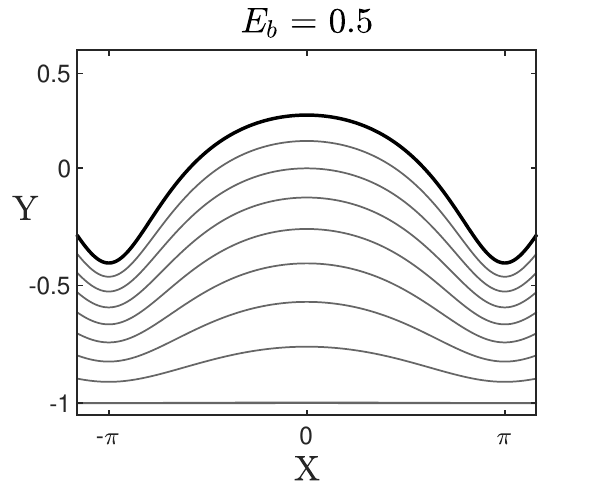}
		\includegraphics[scale=1]{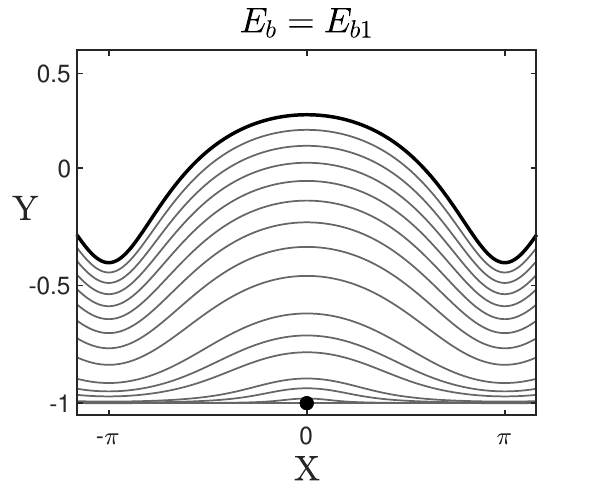}
		\includegraphics[scale=1]{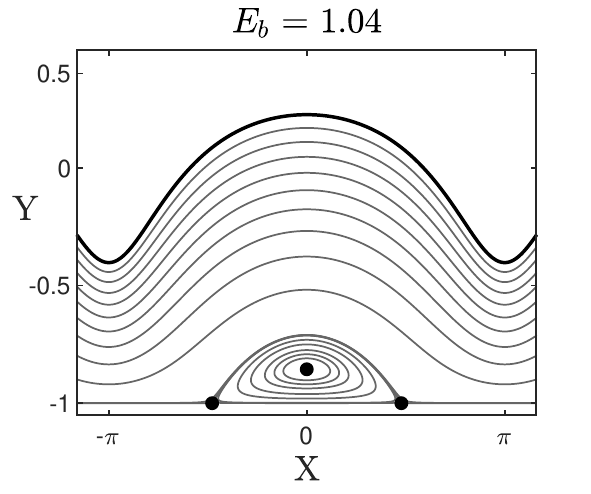}
		\includegraphics[scale=1]{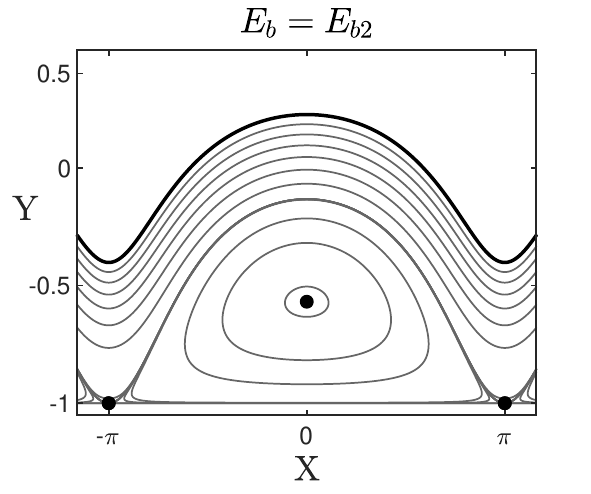}
		\includegraphics[scale=1]{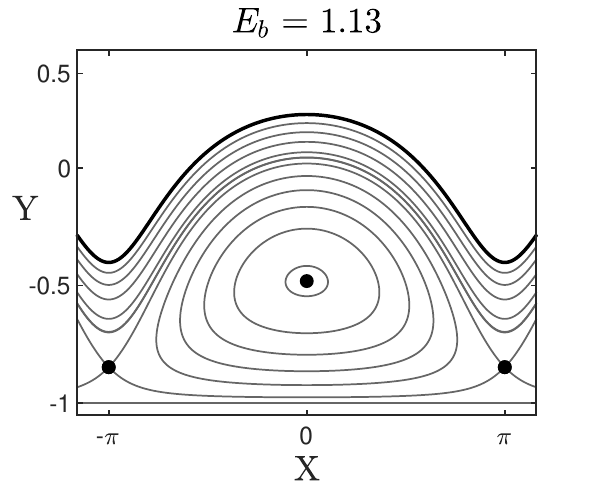}
		
		\caption{Phase portrait for different values of the  electric bond number  $E_b$ ($E_{b1} \approx 0.979$, $E_{b2} \approx 1.118$) Parameters of the flow: $\tau = 1$ and $\gamma = 0.5$. }
		\label{fig:PhasePortraitTau_1}
	\end{figure}

	\subsection {The electric Bond number $E_b$  and the appearance of stagnation points}

	We start by analysing  how the electric field strength  affects  the appearance and the location of  stagnation points. For this purpose, we fix the vorticity and the surface tension by setting $\gamma = 0.5$ and $\tau = 0$,  and let $E_b$ vary. 	In figure  \ref{fig:PhasePortraitTau_0}, a  parameter sweep is shown for values of $E_b = 0$ up to $E_b = 0.88$. 
	In the absence of an electric field, i.e., when $E_b = 0$,  there is no stagnation point in the flow. However, a gradual increase in $E_b$ leads to the emergence of a single stagnation point on the bottom boundary below the crest. This bifurcation occurs at a specific value $E_{b1}$ which is shown to be approximately $0.638$ by the numerical results (see the middle left panel of Figure \ref{fig:PhasePortraitTau_0}).  Once $E_b$ crosses this threshold value $E_{b1}$, the single stagnation point splits into two saddles located on the bottom and one centre located in the bulk of the fluid below the crest, forming a recirculation zone attached to the bottom boundary as displayed in the middle right panel of Figure \ref{fig:PhasePortraitTau_0}. As $E_b$ is further increased, the centre moves vertically towards the surface. Meanwhile, the saddle points on the bottom  move horizontally apart  from each other towards the two periodic boundaries at $X=\pm \pi$ respectively. They eventually collide with the saddle points from the neighbourhood periods  right below the crest when  $E_b$ is equal to a second threshold value denoted by $E_{b2}$ that is about $0.803$. As a result,  the flow structure beneath the surface, as shown in the bottom left panel of Figure \ref{fig:PhasePortraitTau_0}, has a centre located below the crest in the bulk of the fluid and one saddle located below the trough on the bottom boundary. By increasing  further the value of $E_b$, the saddle point detaches from the bottom and  moves upwards, which gives rise to a recirculation zone whose streamlines are characterized by the well known  Kelvin cat-eye pattern  -- see the bottom right panel of Figure \ref{fig:PhasePortraitTau_0}.    Looking at this figure one may reckon that a slight increasing in the value of $E_b$  can lead to  the stagnation points reach the free surface. However, the value $E_b = 0.88$ is the maximum one for which our numerical method captures a wave solution. Therefore, we cannot conclude whether such behaviour occurs or not.  Nonetheless, it is worth to mention that in the linear case this not happen. This is easily seen from equations  (\ref{eq:delta}) and (\ref{eq:TDE1}).

	In the presence of surface tension, the wave profile possess broader crests and sharper troughs comparing to the case without capillarity at same amplitude. Consequently, the flow structure beneath the surface adopts the feature as shown in the top panels in Figure \ref{fig:PhasePortraitTau_1} for the case where $\tau=1$. By increasing the value of $E_b$, qualitatively similar results have been obtained.  Again, the evolution of the stagnation points can be briefly summarised as follows.
	\begin{enumerate}
	\item No stagnation points exist for $E_b<E_{b1}$ in which the critical value $E_{b1}$ is observed to be about  $0.979$.
	    \item The appearance of a stagnation point takes place at $E_{b1}$.
	    \item The stagnation point splits into three: a centre shifting upwards and two saddle points on the bottom boundary drifting away from each other with further increase of $E_b$ until the second critical value $E_{b2}$ that is approximately $1.118$.
	    \item The saddle points collide with the ones from the neighbourhood periods at $X=\pm \pi$ when $E_b=E_{b2}$. The stagnation points now reduce to a centre and a saddle point within one wavelength. 
	    \item The saddle point and the centre both move up  vertically towards the free-surface until the upper limit of $E_b$ is attained for the stable regime. 
	\end{enumerate}
	It is observed that the threshold values $E_{b1}$ and  $E_{b2}$ become greater as the surface tension gets stronger. This feature indicates that stronger electric fields are required for the formation of one or more stagnation points against the surface tension. The surface tension also results in a larger recirculation zone as presented in Figure \ref{fig:PhasePortraitTau_1} in comparison to Figure \ref{fig:PhasePortraitTau_0}, and the corners of the cat-eye being close to the bottom boundary even in the presence of strong normal electric fields.

	\subsection{Pressure anomalies}
	It is well known that the maximum pressure value is attained on the bottom below the crest \cite{CS10} in the context of gravity irrotational flows. However, this feature is not necessarily true for flows with constant vorticity  \cite{VO}. The work done by Ribeiro-Jr \emph{et al.} \cite{ribeiro2017flow} has inferred that  such pressure anomalies are related to the emergence of saddle points. A step forward on this topic is to understand the role of normal electric fields in such anomalies. 
	\begin{figure}[!ht]
		\centering
		\includegraphics[scale=1]{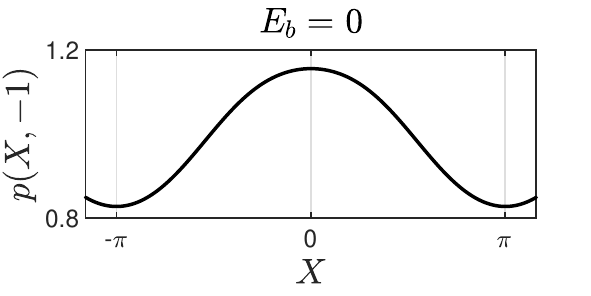}
		\includegraphics[scale=1]{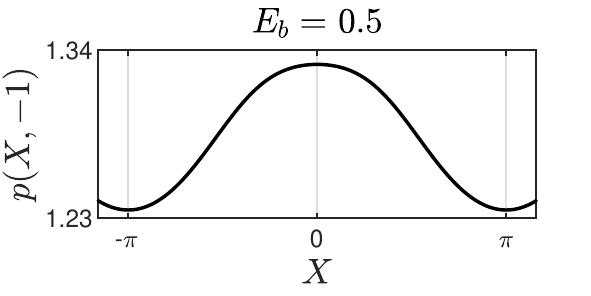}
		\includegraphics[scale=1]{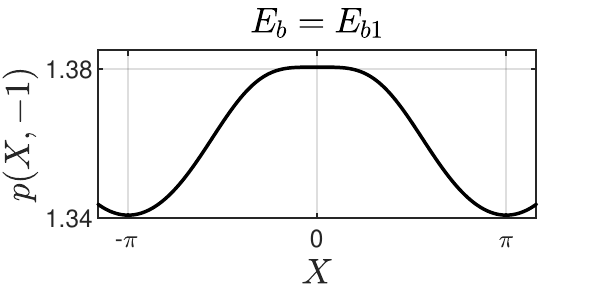}
		\includegraphics[scale=1]{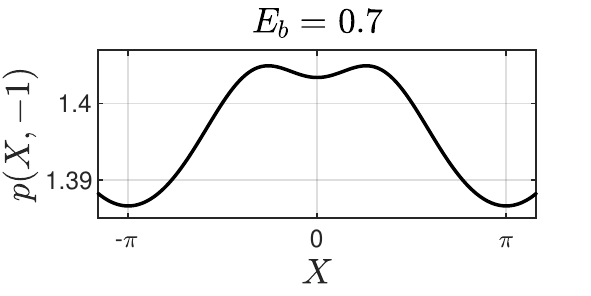}
		\includegraphics[scale=1]{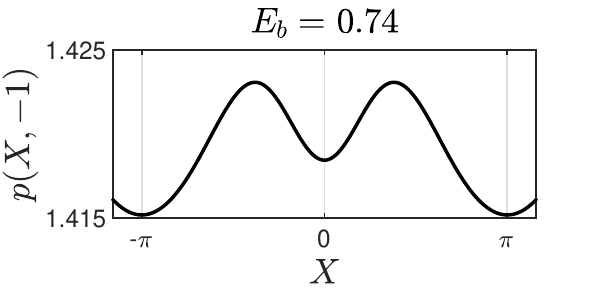}
		\includegraphics[scale=1]{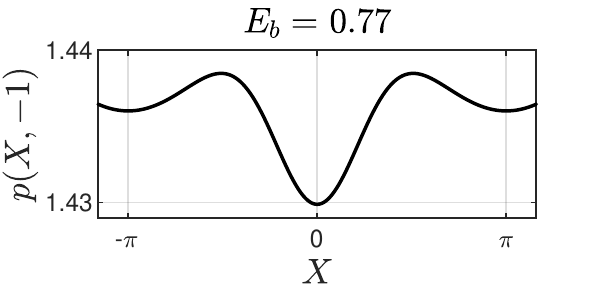}
		\includegraphics[scale=1]{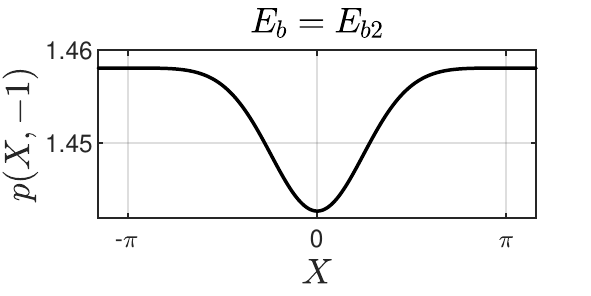}
		\includegraphics[scale=1]{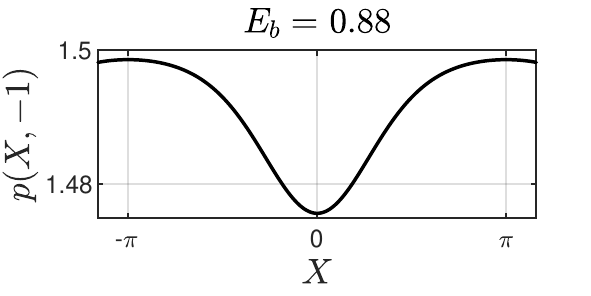}
		\caption{Pressure on the bottom  for different values of $E_b$.
			The free surface wave corresponding to each pressure is depicted in Figure \ref{fig:PhasePortraitTau_0}, except for $E_b = 0.74$ and $E_b = 0.77$.}
		\label{fig:Pressure_on_the_bottom_Tau_zero}
	\end{figure}

\begin{figure}[!ht]
		\centering
		\includegraphics[scale=1]{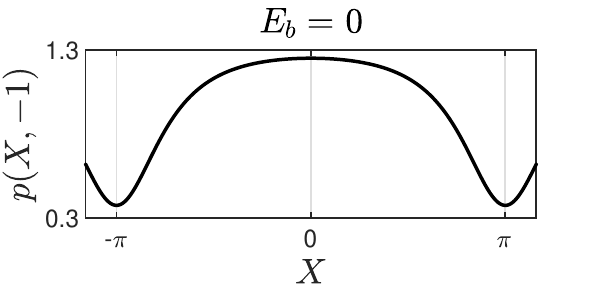}
		\includegraphics[scale=1]{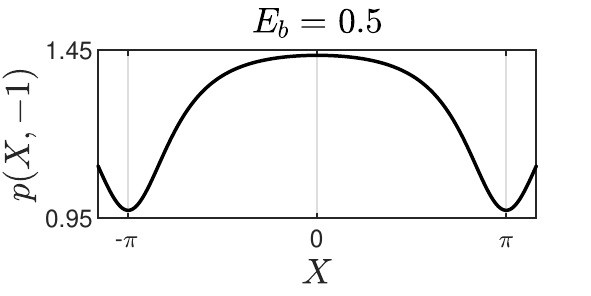}
		\includegraphics[scale=1]{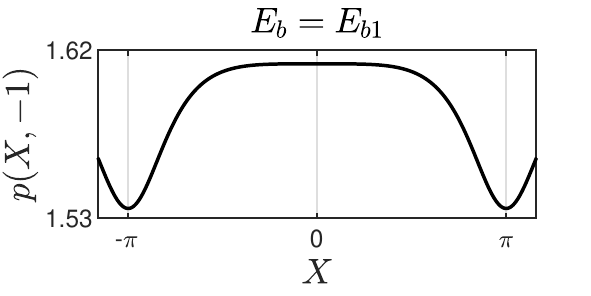}
		\includegraphics[scale=1]{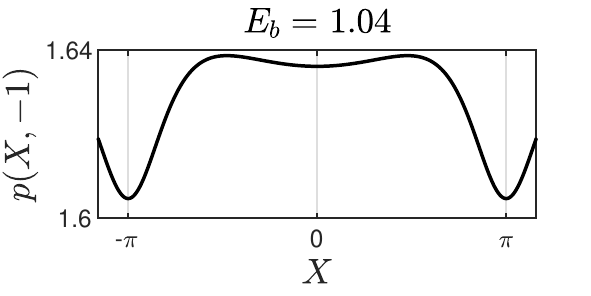}
		\includegraphics[scale=1]{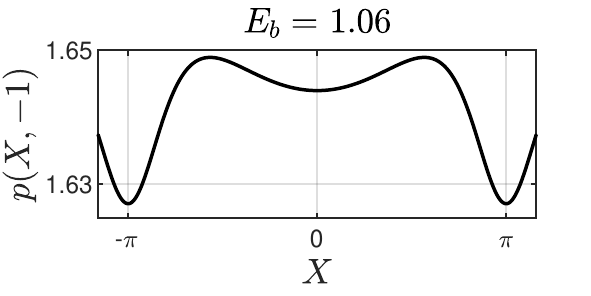}
		\includegraphics[scale=1]{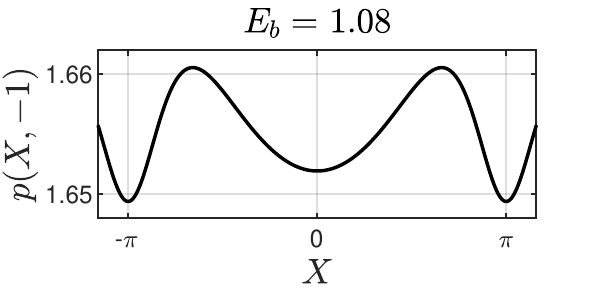}
		\includegraphics[scale=1]{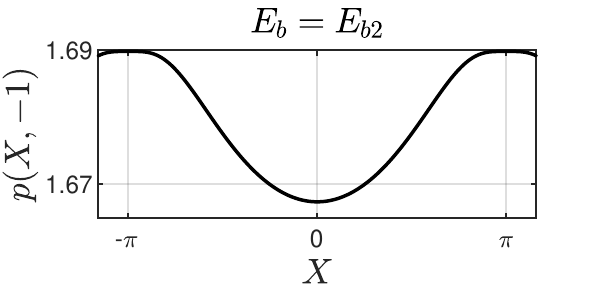}
		\includegraphics[scale=1]{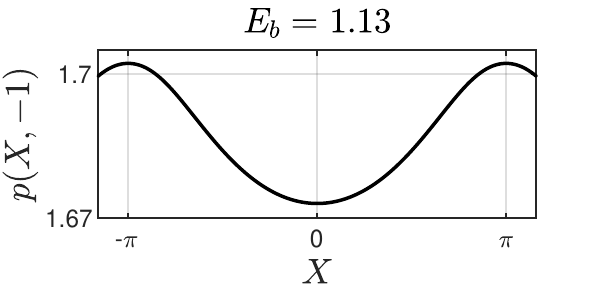}
		\caption{Pressure on the bottom  for different values of $E_b$.
			The free surface wave corresponding to each pressure is depicted in Figure \ref{fig:PhasePortraitTau_1}, except for $E_b = 1.06$ and $E_b = 1.08$.}
		\label{fig:Pressure_on_the_bottom_Tau_one}
	\end{figure}

The pressure on the bottom boundary is presented in figure \ref{fig:Pressure_on_the_bottom_Tau_zero}  for the same parameters values  ($\tau = 0$ and $\gamma = 0.5$) used in figure \ref{fig:PhasePortraitTau_0}.	In the absence of electric fields, the pressure on the bottom boundary is characterized by attaining the local maxima and  minima beneath the crest and the trough of the free surface wave, respectively. However, with the increase of  $E_b$,  the pressure on the bottom experience a series of changes until it reaches a configuration in which the  local maxima and the local minima are located  below the trough and  the crest, respectively.  Qualitatively similar behaviour is found in the presence of surface tension for  capillary-gravity flows with constant vorticity, see figure \ref{fig:Pressure_on_the_bottom_Tau_one} for an example with $\tau=1$  and $\gamma=0.5$. The finding is of great interest to the engineering community since it demonstrates that  normal electric fields are capable of modifying significantly the hydrodynamic pressure beneath a fluid surface with only limited change to the wave profile.

\begin{figure}[!h]
		\centering
		\includegraphics[scale=1]{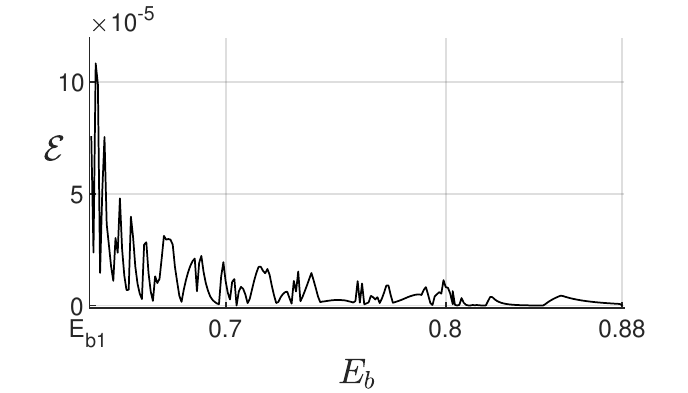}
		\caption{Comparison between the abscissa of  maximum point  of the pressure on the bottom  $(x_b)$ and the abscissa  of the saddle point $(x_s)$ for different values of $E_b$. $\mathcal{E} =  \dfrac{ |x_b-x_s|}{|x_s|}$. Parameters of the flow: $\tau = 0$ and $\gamma = 0.5$}
		\label{fig:Comparsion_tau_zero}
	\end{figure}

\begin{figure}[!h]
		\centering
		\includegraphics[scale=1]{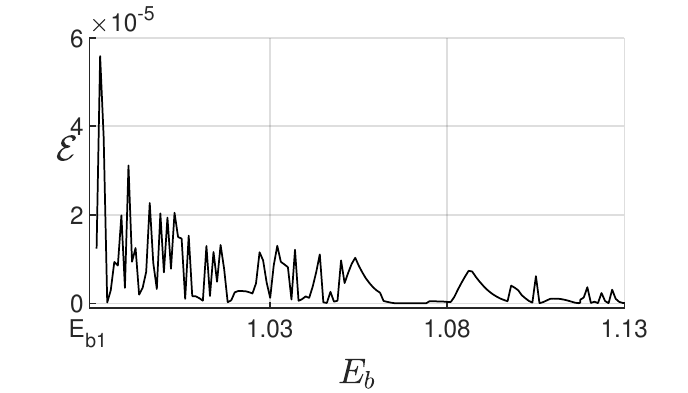}
		\caption{Comparison between the abscissa of  maximum point  of the pressure on the bottom  $(x_b)$ and the abscissa  of the saddle point $(x_s)$ for different values of $E_b$. $\mathcal{E} =  \dfrac{ |x_b-x_s|}{|x_s|}$.  Parameters of the flow: $\tau = 1$ and $\gamma = 0.5$.}
		\label{fig:Comparsion_tau_one}
	\end{figure}

	In order to examine whether the pressure anomalies are caused by the formation of saddle points for the present problem,  we compute
	\begin{equation}\label{eq:Error}
		\mathcal{E}(E_b) =  \frac{ |x_p(E_b)-x_s(E_b)|}{|x_s(E_b)|},
	\end{equation}
	for various $E_b$. Here $x_p(E_b)$ is the abscissa of the local maximum of the pressure on the bottom boundary and $x_s(E_b)$  is the abscissa of the saddle point.  This quantity $\mathcal{E}$ displays the relative distance between the saddle point and the local maximum of the pressure and is found to be very close to zero regardless of the value of $\tau$. Two examples are depicted in Figure  \ref{fig:Comparsion_tau_zero} and \ref{fig:Comparsion_tau_one} for $\gamma = 0.5$,   $\tau =0$, and  $\gamma = 0.5$,  $\tau = 1$, respectively. Hence, we discover a strong numerical evidence that $x_p$ and $x_s$ are identical and $\mathcal{E}$ from \eqref{eq:Error} is purely due to numerical errors.	It is concluded that the  projections of saddle points onto the bottom boundary correspond to regions of high pressure.

	%

	\section{Conclusion}\label{conclusion}
	In this paper, we  performed a quantitative study on the effects	of normal electric fields in the flow structure beneath  periodic capillary-gravity travelling waves with constant vorticity. By fixing the vorticity, it was discovered that the flow can have zero, one, two or three stagnation points depending on the value of $E_b$. The appearance of a single stagnation point was found to occur at $E_b=E_{b1}$.	When $E_b>E_{b1}$, there exist multiple stagnation points which form a  recirculation zone. At $E_b=E_{b2}$, two saddle points  from consecutive periods collide and merge into one which shifts away from the bottom boundary with a further increase of $E_b$. More interestingly,  the impact of the electric effect on the pressure beneath the surface was shown to be significant. 	Besides, we demonstrated that the locations of  local maxima for the pressure on the lower boundary are intrinsically associated to the projections of the saddle points onto the bottom. To the best of our knowledge, the study carried out in this paper is the first one in the literature devoted to the investigation of stagnation points and pressure anomalies in the context of electrohydrodynamics  flows. It opens the door to other fronts of study in the such field which eventually can lead to investigating more realistic flows.

	\section*{Acknowledgments}
	 R.R.-Jr is grateful to University
	of Bath for the extended visit to the Department of Mathematical Sciences. T. G. was supported by the Numerical and Applied Mathematics Research Unit, School of Computing and Mathematical Sciences, University of Greenwich.

	\section*{Data Availability Statement}
	Data sharing is not applicable to this article as the parameters used in the numerical experiments are informed in this paper.

	\end{document}